\documentclass [fleqn,12pt]{article} 
\onecolumn         
\title{
\noindent {\small BULLETIN DE L'ACADEMIE POLONAISE DES SCIENCES} \\
\noindent {\small Serie des sciences math., ast.  et phys. -- Vol. XXV, No. 5, 
p. 521, 1977} \\
\noindent {\small \underline{THEORETICAL~PHYSICS}}  \\
\vspace{0.3cm}
Irreducible Tensor Operators and the Wigner-Eckart Theorem  for
  Finite Magnetic Groups \\ \vspace{1.0cm} by} 
\author{Prasanta RUDRA \\  \vspace{1.0cm} {\sl Presented by J. RZEWUSKI
  on June 21, 1976}}
\date{ }
\begin{document}
\maketitle
\vspace{1.0cm}
{\bf Summary.} The transformation properties of irreducible tensor operators
and the applicability of the Wigner-Eckart theorem to finite magnetic
groups have been studied.
\section{Introduction} \label{intro}
Selection rules and ratio of intensities for
transitions between different states of a physical system are obtained
from the appropriate matrix elements of operators between the initial
and the final states of the system \cite{rose, judd}. The calculation
of the matrix elements becomes simplified if one invokes the Wigner-Eckart
theorem \cite{wigner, eckart, fano}. This theorem introduces the
concept of a set of operators that transforms according to some
irreducible representation of the appropriate symmetry group of the
system. For compact and for finite groups, if the Kronecker inner
direct product of two irreducible representations contains any irreducible
representation only once, the matrix element of such an operator between
states belonging to irreducible representations will be proportional to
the corresponding Clebsch-Gordan (CG) coefficient, the proportionality
constant being called the reduced matrix element. The proof of this
theorem depends on the fact that the matrix element when transformed
by a symmetry element of the  system has the same value as the
untransformed matrix element. If the  symmetry group of the system is a
magnetic group, then antilinear elements \cite{wigner} are present and
for these elements the transformed matrix elements are complex conjugate
of the untransformed value. For this reason the Wigner-Eckart theorem is
not in general valid in the case of magnetic groups. Recently Backhouse
\cite{backhouse} and Doni and Paravicini \cite{doni} have investigated
the theory of selection rules in magnetic crystals whose symmetry group
contains antilinear elements. We have here investigated the conditions
when the Wigner-Eckart theorem is valid for symmetry groups containing
antilinear elements. To this end we have studied the transformation
laws of irreducible tensor operators (both linear and antilinear) for
magnetic groups. In order that the results can be applied to spinor
cases as well, projective corepresentations \cite{janssen, bradley} have
been considered. Previously Aviran and Zak \cite{zak} have investigated
this problem. Their results are somewhat complicated because a
quadratic relationship between the CG coefficients was used in their 
analysis, whereas a linear relationship has been used here.
\section{Irreducible tensor operators}  \label{tensor}
Here we give transformation laws of irreducible tensor operators for a
magnetic group \cite{janssen, bradley}
\begin{equation}
M(G) = G \cup a_0 G, ~~~a_0^2\in G,   \label{ito1}
\end{equation}
where $a_0$ is an antilinear element and $G$ is a group of linear elements.
The corepresentation $D^{\lambda}\left(\alpha\right),~~\alpha \in M\left(
 G\right)$, belonging to the cofactor system $\lambda\left(\alpha,\beta
\right),~~\alpha,\beta\in M\left(G\right)$ satisfies \cite{janssen, bradley,
rudra1}
\begin{eqnarray}
D^{\lambda}\left(\alpha\right)D^{\lambda}\left(\beta\right)^{\left[\alpha
\right]} & = & \lambda\left(\alpha,\beta\right)^{\left[\alpha\beta\right]}
 D^{\lambda}\left(\alpha\beta\right),  \nonumber \\
\lambda\left(\alpha,\beta\right)^{\left[\gamma\right]}\lambda\left(
\alpha\beta,\gamma\right) & = & \lambda\left(\alpha,\beta\gamma\right)
 \lambda\left(\beta,\gamma\right),  \label{ito2}  \\
|\lambda\left(\alpha,\beta\right)| & = & 1.  \nonumber 
\end{eqnarray}
We have used the square bracket symbol $\left[\alpha\right]$ everywhere,
so that
\begin{displaymath}
A^{\left[\alpha\right]} = \left\{\begin{array}{l} A,~{\rm if}~\alpha~
 {\rm is~linear}, \\ A^{\ast},~{\rm if}~\alpha~{\rm is~antilinear,}
 \end{array} \right. 
\end{displaymath}
where $A$ is a matrix, an operator, or a complex number.

We define the Wigner operator \cite{wigner} $O_{\alpha},~\alpha\in
M\left(G\right)$, by the relation
\begin{equation}
O_{\alpha}O_{\beta}^{\left[\alpha\right]} = \lambda\left(\alpha,\beta
\right)^{\left[\alpha\beta\right]}O_{\alpha\beta}.  \label{ito3}
\end{equation}
This relation is satisfied when $O_{\alpha}$s operate on the bases
belonging to the appropriate cofactor system. For proper rotations
characterized by the Eulerian angles $\left(\alpha, \beta, \gamma\right)$
\begin{equation}
O_{\left(\alpha,\beta,\gamma\right)}=\exp\left(-i\alpha J_z\right)
\exp\left(-i\beta J_y\right)\exp\left(-i\gamma J_z\right), \label{ito4}
\end{equation}
where $J_i$s are the usual angular momentum operators. The relation
(\ref{ito3}) will be automatically satisfied if we take the
appropriate bases belonging to the vector corepresentation or the
spinor corepresentation. For the time reversal operator $\theta$, which
is antilinear, its action on the spin states $|j,m\rangle$ will be
given by
\begin{equation}
O_{\theta}|j,m\rangle = \left(-1\right)^{j-m} |j,-m\rangle.  \label{ito5}
\end{equation}
The $m$-th component of any tensor operator belonging to the $\mu$-th
irreducible corepresentation of the cofactor system $\lambda\left(\alpha,
\beta\right)$, which may be either linear or antilinear, will transform
as
\begin{equation}
T_m^{\lambda\mu}\left(\alpha\right) = \lambda\left(\alpha^{-1},\alpha
\right)^{\ast\left[T\right]}O_{\alpha}T_m^{\lambda\mu}O_{\alpha^{-1}
 }^{\left[\alpha\right]} = \sum_n D_{nm}^{\lambda\mu}\left(\alpha
 \right)^{\ast\left[T\right]}T_n^{\lambda\mu}.   \label{ito6}
\end{equation}
This relation will also cover the case when $T$ is antilinear. For the
sake of completeness we write here the transformation relation of the
bases belonging to the $\mu$-th irreducible corepresentation of the
same factor system $\lambda\left(\alpha,\beta\right)$
\begin{displaymath}
\left|\right.\Phi_m^{\lambda\mu}\left(\alpha\right)\left.\right\rangle = 
O_{\alpha} \left|\right.\Phi_m^{\lambda\mu}\left.\right\rangle = \sum_n
D_{nm}^{\lambda\mu}\left(\alpha\right)\left|\right.\Phi_n^{\lambda\mu}
\left. \right\rangle.  \hspace{5.5cm} \left(6{\rm a}\right)
\end{displaymath}
Thus
\begin{displaymath}
T_m^{\lambda\mu}\left(\alpha\right)\left|\right.\Phi\left(\alpha\right)
\left.\right\rangle = O_{\alpha}T_m^{\lambda\mu}\left|\right.\Phi
\left.\right\rangle.  \hspace{7.8cm} \left(6{\rm b}\right)
\end{displaymath}

The result of successive action of two operators $O_{\beta}$ and 
$O_{\alpha}$ will be given by
\begin{equation}
O_{\alpha}O_{\beta}T_m^{\lambda\mu}O_{\beta^{-1}}^{\left[\beta\right]}
O_{\alpha^{-1}}^{\left[\alpha\right]} = \lambda\left(\alpha,\beta
\right)^{\left[\alpha\beta\right]}\frac{\lambda\left(\alpha^{-1},\alpha
\right)\lambda\left(\beta^{-1},\beta\right)^{\left[\alpha\right]}}{\lambda
\left(\beta^{-1}\alpha^{-1},\alpha\beta\right)}\cdot O_{\alpha\beta}
T_m^{\lambda\mu}O_{\beta^{-1}\alpha^{-1}}^{\left[\alpha\beta\right]}.
\label{ito7}
\end{equation}
The proof is a straightforward application of Eq.~\ref{ito2} forthe
choice $\lambda\left(\alpha,e\right)=\lambda\left(e,\alpha\right)=1$.

The irreducible tensors $T_m^{\lambda\mu}$ can be obtained by the
operation of the projection operator $P_m^{\lambda\mu}$ on an
arbitrary tensor $T$
\begin{equation}
T_m^{\lambda\mu} = P_m^{\lambda\mu}T = \sum_{\alpha\in M} 
D_{mm_0}^{\lambda\mu}\left(\alpha\right)^{\left[T\right]}\lambda\left(
\alpha^{-1},\alpha\right)^{\ast\left[T\right]} O_{\alpha}T
O_{\alpha^{-1}}^{\left[\alpha\right]},   \label{ito8}
\end{equation}
where $m_0$ is any fixed index.

Incidentally, the projection operator $P_m^{\lambda\mu}$, which
operating on an arbitrary state $\left|\right.\Phi\left.\right\rangle$ will
give the $m$-th basis of the $\mu$-th irreducible corepresentation
belonging to the cofactor system $\lambda\left(\alpha,\beta\right)$ has
the same form \cite{rudra2} as for the vector corepresentation.
\begin{equation}
P_m^{\lambda\mu} = \sum_{\alpha\in M}D_{mm_0}^{\lambda\mu}\left(\alpha
\right)^{\ast}O_{\alpha}.    \label{ito9}
\end{equation} 
\section{Wigner-Eckart theorem}  \label{wet}
For groups with linear elements the Wigne-Eckart theorem \cite{wigner}
states that under the restriction given in Sec.~\ref{intro}
\begin{equation}
\left\langle\Phi_{m_3}^{\lambda_1\mu_1}\left|T_{m_2}^{\lambda_2\mu_2}
\right|\Phi_{m_3}^{\lambda_3\mu_3}\right\rangle = \frac{1}{d_{\lambda_3
\mu_3}}\left\langle\lambda_1 \mu_1\left|\left|\lambda_2\mu_2\right|\right|
\lambda_3\mu_3\right\rangle\left\langle\lambda_1\mu_1 m_1;\lambda_2\mu_2
m_2\left|\right.\lambda_3\mu_3 m_3\right\rangle,  \label{wet10}
\end{equation}
where $\left\langle\lambda_1\mu_1 m_1;\lambda_2\mu_2 m_2\left | \right.
\lambda_3\mu_3 m_3\right\rangle$ is the CG coefficient defined by the 
relation
\begin{equation}
\left|\right.\Phi_{m_3}^{\lambda_3\mu_3}\left.\right\rangle = \sum_{m_1m_2}
\left\langle\lambda_1\mu_1 m_1;\lambda_2\mu_2 m_2\left|\right.\lambda_3
\mu_3 m_3\right\rangle\left|\right.\Phi_{m_1}^{\lambda_1\mu_1}\left.
\right\rangle \left.\right|\Phi_{m_2}^{\lambda_2\mu_2}\left.\right\rangle.
\label{wet11}
\end{equation}
The reduced matrix element is defined by
\begin{equation}
\left\langle\lambda_1\mu_1\left|\right|\lambda_2\mu_2\left|\right|
\lambda_3\mu_3\right\rangle = \sum_{n_1n_2n_3}
\left\langle\lambda_1\mu_1n_1;\lambda_2\mu_2 n_2\left.\right|\lambda_3
\mu_3 n_3\right\rangle^{\ast}\left\langle\Phi_{n_1}^{\lambda_1\mu_1}
\left|T_{n_2}^{\lambda_2\mu_2}\right|\Phi_{n_3}^{\lambda_3\mu_3}\right\rangle
\label{wet12}
\end{equation}
and $d_{\lambda_3\mu_3}$ = the dimension of the irreducible corepresentation
$D^{\lambda_3\mu_3}$.

The CG coefficient is zero unless $\lambda_3\left(\alpha,\beta\right) =
\lambda_1\left(\alpha,\beta\right)\lambda_2\left(\alpha,\beta\right)$.

For magnetic groups no such simple relation is, in general, true. We now
investigate the conditions for the validity of such a theorem for
magnetic groups. We note that
\begin{eqnarray}
\left\langle\Phi_{m_1}^{\lambda_1\mu_1}\left(\alpha\right)\left|
T_{m_2}^{\lambda_2\mu_2}\left(\alpha\right)\right|\Phi_{m_3}^{\lambda_3\mu_3}
\left(\alpha\right)\right\rangle & = & \sum_{n_1n_2n_3}\left\langle
\Phi_{n_1}^{\lambda_1\mu_1}\left|T_{n_2}^{\lambda_2\mu_2}\right|
\Phi_{n_3}^{\lambda_3\mu_3}\right\rangle\times   \nonumber  \\
 &   & \hspace{0.3cm} D_{n_1m_1}^{\lambda_1\mu_1}\left(\alpha\right)^{\ast}
 D_{n_2m_2}^{\lambda_2\mu_2}\left(\alpha\right)^{\ast\left[T\right]}
 D_{n_3m_3}^{\lambda_3\mu_3}\left(\alpha\right)^{\left[T\right]}.
\label{wet13}
\end{eqnarray}

Case 1. $T$ is a linear operator.  \\

In this case the matrix element on the left-hand side of Eq.~(\ref{wet13})
is zero unless
\begin{equation}
\lambda_3\left(\alpha,\beta\right) = \lambda_1\left(\alpha,\beta\right)
\lambda_2\left(\alpha,\beta\right),~~~~\forall \alpha,\beta\in M\left(
G\right).   \label{wet14}
\end{equation}
Using the transformation relation (\ref{ito6}) and (6a) for 
$T_m^{\lambda\mu}\left(\alpha\right)$ and $\Phi_m^{\lambda\mu}\left(
\alpha\right)$ and the linear equations (Eq.~(27) of Ref.~\cite{rudra1})
satisfied by the CG coefficients of a magnetic group
\begin{eqnarray}
\sum_{m_1m_2}\left[\left\langle\lambda_1\mu_1 m_1;\lambda_2\mu_2\left.
\right|\lambda_3\mu_3 m_3\right\rangle\sum_{u\in G}
D_{i_1m_1}^{\lambda_1 \mu_1}\left(u\right)D_{i_2m_2}^{\lambda_2\mu_2}
\left(u\right)D_{i_3m_3^{\prime}}^{\lambda_3\mu_3}\left(u\right)^{\ast} 
\right. ~~~~~&  & \nonumber \\
~~~~+\left.\left\langle\lambda_1\mu_1 m_1;\lambda_2\mu_2 m_2
\left.\right| \lambda_3\mu_3 m_3^{\prime}\right\rangle^{\ast}\sum_{a\in M-G}
D_{i_1m_1}^{\lambda_1 \mu_1}\left(a\right)D_{i_2m_2}^{\lambda_2\mu_2}
\left(a\right)D_{i_3m_3}^{\lambda_3\mu_3}\left(a\right)^{\ast} 
\right]   &   &   \nonumber  \\
~~~~=~\frac{\left|M\right|}{d_{\lambda_3\mu_3}}\delta_{m_3,m_3^{\prime}}
\left\langle\lambda_1\mu_1i_1;\lambda_2\mu_2i_2\left.\right|\lambda_3
\mu_3i_3\right\rangle,  &   &   \label{wet15}
\end{eqnarray}
we get
\begin{eqnarray}
\sum_{m_1m_2}
\left[\left\langle\lambda_1\mu_1 m_1;\lambda_2\mu_2m_2\left.
\right|\lambda_3\mu_3 m_3^{\prime}\right\rangle^{\ast}\sum_{u\in G}
\left\langle\Phi_{m_1}^{\lambda_1 \mu_1}\left(u\right)\left|
T_{m_2}^{\lambda_2\mu_2} \left(u\right)\right|\Phi_{m_3}^{\lambda_3\mu_3}
\left(u\right)\right\rangle 
\right. ~~~~~&  & \nonumber \\
~~~~+ \left.\left\langle\lambda_1\mu_1 m_1;\lambda_2\mu_2m_2\left.
\right|\lambda_3\mu_3 m_3\right\rangle\sum_{a\in M-G}
\left\langle\Phi_{m_1}^{\lambda_1 \mu_1}\left(a\right)\left|
 T_{m_2}^{\lambda_2\mu_2} \left(a\right)\right|
\Phi_{m_3^{\prime}}^{\lambda_3\mu_3} \left(a\right)\right\rangle 
\right]   &   &   \nonumber  \\
~~~~=~\frac{\left|M\right|}{d_{\lambda_3\mu_3}}\delta_{m_3,m_3^{\prime}}
\sum_{n_1n_2n_3}\left\langle\lambda_1\mu_1n_1;\lambda_2\mu_2n_2\left.
\right|\lambda_3 \mu_3n_3\right\rangle^{\ast}\left\langle
\Phi_{n_1}^{\lambda_1\mu_1}\left|T_{n_2}^{\lambda_2\mu_2}\right|
\Phi_{n_3}^{\lambda_3\mu_3}\right\rangle,  &   &   \label{wet16}
\end{eqnarray}
where $\left|M\right| = {\rm the~order~of~the~magnetic~group}~M\left(G
\right)$.

In expressing
\begin{displaymath}
\left\langle\Phi_{m_1}^{\lambda_1\mu_1}\left(\alpha\right)\left|
T_{m_2}^{\lambda_2\mu_2}\left(\alpha\right)\right|\Phi_{m_3}^{\lambda_3
\mu_3}\left(\alpha\right)\right\rangle
\end{displaymath}
in Eq.~\ref{wet16} we shall use the identity \cite{rudra1}
\begin{displaymath}
\left\langle O_{\alpha_1}z_1\Phi_1\left|\right.O_{\alpha_2}z_2\Phi_2
\right\rangle = z_1^{\ast\left[\alpha_1\right]}z_2^{\left[\alpha_2\right]}
\left\langle\Phi_1\left|O_{\alpha_1^{-1}\alpha_2}\right|\Phi
\right\rangle^{\left\{\alpha_1,\alpha_2\right\}}
\end{displaymath}
with
\begin{equation}
\left\langle\Phi_1\left|O_{\alpha_1^{-1}\alpha_2}\right|\Phi_2
\right\rangle^{\left\{\alpha_1,\alpha_2\right\}} = \left\{\begin{array}{l}
\left\langle\Phi_2\left|O_{\alpha_2^{-1}\alpha_1}\right|\Phi_1\right\rangle
~~~{\rm if~both}~\alpha_1,\alpha_2\in M-G \\
\left\langle\Phi_1\left|O_{\alpha_1^{-1}\alpha_2}
\right|\Phi_2\right\rangle^{\left[\alpha_1\right]}
~~~{\rm otherwise}\end{array} \right.     \label{wet17}
\end{equation}
where $\Phi_i$s are state vectors and $z_i$s are complex numbers. We
observe that the expression on the left-hand side of Eq.~(\ref{wet16}) is
real and we can write
\begin{eqnarray}
\frac{1}{2}\sum_{m_1m_2}
\left[\left\langle\lambda_1\mu_1 m_1;\lambda_2\mu_2m_2\left.
\right|\lambda_3\mu_3 m_3^{\prime}\right\rangle^{\ast}
\left\langle\Phi_{m_1}^{\lambda_1 \mu_1}\left|
T_{m_2}^{\lambda_2\mu_2} \right|\Phi_{m_3}^{\lambda_3\mu_3}
\right\rangle\right. ~~~~~&  & \nonumber \\
~~~~+ \left.\left\langle\lambda_1\mu_1 m_1;\lambda_2\mu_2m_2\left.
\right|\lambda_3\mu_3 m_3\right\rangle
\left\langle\Phi_{m_1}^{\lambda_1 \mu_1}\left|
 T_{m_2}^{\lambda_2\mu_2} \right|
\Phi_{m_3^{\prime}}^{\lambda_3\mu_3} \right\rangle 
\right]   &   &   \nonumber  \\
~~~~=~\frac{1}{d_{\lambda_3\mu_3}}\delta_{m_3,m_3^{\prime}}
\left\langle\lambda_1\mu_1\left|\left|\lambda_2\mu_2\right|
\right|\lambda_3 \mu_3\right\rangle_L,  &   & \hspace{2.0cm}  (17a) \nonumber 
\end{eqnarray}
where for a linear operator $T$ the reduced matrix element is given by
\begin{equation}
\left\langle\lambda_1\mu_1\left|\left|\lambda_2\mu_2\right|\right|
\lambda_3\mu_3\right\rangle_L = \sum_{n_1n_2n_3}\left\langle\lambda_1
\mu_1n_1;\lambda_2\mu_2n_2\left|\right.\lambda_3\mu_3n_3
\right\rangle^{\ast}\left\langle\Phi_{n_1}^{\lambda_1\mu_1}\left|
T_{n_2}^{\lambda_2\mu_2}\right|\Phi_{n_3}^{\lambda_3\mu_3}\right\rangle.
    \label{wet18}
\end{equation}

These will be the equations satisfied by the matrix elements. The CG
coefficients satisfy the orthogonality relations \cite{rudra1}
\begin{eqnarray}
\sum_{m_1m_2m_1^{\prime}m_2^{\prime}}\left\langle\lambda_1\mu_1m_1;
\lambda_2\mu_2m_2\left|\right.\lambda_3\mu_3m_3\right\rangle^{\ast}
\left\langle\lambda_1\mu_1m_1^{\prime};\lambda_2\mu_2m_2^{\prime}
\left|\right.\lambda_3\mu_3^{\prime}m_3^{\prime}\right\rangle^{\ast}
  \times  &   &   \nonumber  \\
\left\langle\lambda_1\mu_1m_1^{\prime}\left|\right.\lambda_1\mu_1m_1
\right\rangle  
\left\langle\lambda_2\mu_2m_2^{\prime}\left|\right.\lambda_2\mu_2m_2
\right\rangle 
 = \delta_{\mu_3\mu_3^{\prime}}\left\langle\lambda_3\mu_3m_3^{\prime}
\left|\right.\lambda_3\mu_3m_3\right\rangle &    &  \label{wet19}
\end{eqnarray}
It should be noted that the bases belonging to either a type (a) or a
type (c) corepresentation \cite{wigner} are all orthogonal  \cite{rudra1}.
Thus if none of the 3 corepresntations $\mu_1,\mu_2:\mu_3$ are of
Wigner type (b), then
\begin{equation}
\left\langle\Phi_{m_1}^{\lambda_1\mu_1}\left|T_{m_2}^{\lambda_2\mu_2}
\right|\Phi_{m_3}^{\lambda_3\mu_3}\right\rangle = \frac{1}{d_{\lambda_3
\mu_3}}\left\langle\lambda_1 \mu_1\left|\left|\lambda_2\mu_2\right|\right|
\lambda_3\mu_3\right\rangle_L\left\langle\lambda_1\mu_1 m_1;\lambda_2\mu_2
m_2\left|\right.\lambda_3\mu_3 m_3\right\rangle,  \label{wet20}
\end{equation}
is a solution of Eq.~(\ref{wet18}). If any of the 3 corepresentations
appearing in the matrix element is of type (b), then the corresponding
matrix element is a linear combination of terms proportional to the
CG coefficients. Even when they are valid, Eq.~(\ref{wet20}) will be
unique only if the CG coefficients obtained from Eq.~(\ref{wet15}) are
unique \cite{rudra1}---\cite{sakata}. The CG coefficients are unique
if and only if
\begin{displaymath}
\det \left|L\left(i_1i_2i_3,m_1m_2m_3\right)+A\left(i_1i_2i_3,m_1m_2m_3
\right)-\delta_{i_1m_1}\delta_{i_2m_2}\delta_{i_3m_3}\right| =0
\end{displaymath}
and
\begin{displaymath}
\det \left|L\left(i_1i_2i_3,m_1m_2m_3\right)-A\left(i_1i_2i_3,m_1m_2m_3
\right)-\delta_{i_1m_1}\delta_{i_2m_2}\delta_{i_3m_3}\right| =0
\end{displaymath}
where
\begin{eqnarray}
L\left(i_1i_2i_3,m_1m_2m_3\right) & = & \frac{d_{\lambda_3\mu_3}}{\left|
M\right|}\sum_{u\in G} D_{i_1m_1}^{\lambda_1\mu_1}\left(u\right)
D_{i_2m_2}^{\lambda_2\mu_2}\left(u\right) D_{i_1m_1}^{\lambda_3\mu_3}
\left(u\right)^{\ast}   \nonumber \\
{\rm and} &   &    \nonumber   \\
A\left(i_1i_2i_3,m_1m_2m_3\right) & = & \frac{d_{\lambda_3\mu_3}}{\left|
M\right|}\sum_{a\in M-G} D_{i_1m_1}^{\lambda_1\mu_1}\left(a\right)
D_{i_2m_2}^{\lambda_2\mu_2}\left(a\right) D_{i_1m_1}^{\lambda_3\mu_3}
\left(a\right)^{\ast}.  \label{wet21} 
\end{eqnarray}

In the case of groups having no antilinear operators if we replace the
second summand on the left-hand side of Eq.~(17a) by the first summand
we get the set of linear equations satisfied by the matrix elements.
Since in these cases the bases are all orthogonal and the CG coefficients
are essentially unique Eq.~(\ref{wet20})is an exact relation
\cite{harter, rudra2}.

In the previous analysis we have assumed that in the expansion of the
Qronecer inner direct product of two irreducible corepresentations
$D^{\lambda_1\mu_1}$ and $D^{\lambda_2\mu_2}$ the irreducible
corepresentation $D^{\lambda_3\mu_3}$ occurs only once. When there are
more than one repetition, the different repetitions of $D^{\lambda_3
\mu_3}$ in the decomposition of the inner product representation are
characterized by $\left\langle\lambda_1\mu_1i_1;\lambda_2\mu_2i_2
\left|\right.\tau_3\lambda_3\mu_3i_3\right\rangle$. As has been shown
in \cite{rudra1} the CG coefficients for different $\tau_3$s
satisfy equations similar to Eq.~(\ref{wet15}). So similar 
considerations will be valid for $\left\langle\Phi_{i_1}^{\lambda_1
\mu_1}\left|T_{i_2}^{\lambda_2\mu_2}\right|\Phi_{i_3}^{\tau_3
\lambda_3\mu_3}\right\rangle$.

But in general the most we can tell about a quantum mechanical state
$\left|\right.\Psi_{i_3}^{\lambda_3\mu_3}$ is that it transforms as
the $i-3$-th component of the irreducible corepresentation
$D^{\lambda_3\mu_3}$. In this case
\begin{displaymath}
\left|\right.\Psi_{i_3}^{\lambda_3\mu_3}\left.\right\rangle = 
\sum_{\tau_3}a_{\tau_3} \left|\right.\Phi_{i_3}^{\tau_3\lambda_3\mu_3}
\left.\right\rangle
\end{displaymath}
and
\begin{displaymath}
\left\langle\Phi_{i_3}^{\lambda_1\mu_1}\left|T_{i_2}^{\lambda_2\mu_2}
\right|\Psi_{i_3}^{\lambda_3\mu_3}\right\rangle = \sum_{\tau_3}
a_{\tau_3}\left\langle\Phi_{i_1}^{\lambda_1\mu_1}\left|T_{i_2}^{\lambda_2
\mu_2}\right|\Phi_{i_3}^{\tau_3\lambda_3\mu_3}\right\rangle.
\end{displaymath}
The quantum mechanical matrix element for the transition probability is
thus a linear combination of terms each of which is a product of a reduced 
matrix element $\left\langle\lambda_1\mu_1\left|\left|\lambda_2\mu_2
\right|\right|\tau_3\lambda_3\mu_3\right\rangle$ and the corresponding
CG coefficient $\left\langle\lambda_1\mu_1i_1;\lambda_2\mu_2\left|\right.
\tau_3\lambda_3\mu_3i_3\right\rangle$.

Case 2. $T$ is an antilinear operator.

In this case the matrix element on the left-hand side of Eq.~(\ref{wet13})
is zero unless
\begin{equation}
\lambda_2\left(\alpha,\beta\right) = \lambda_1\left(\alpha,\beta\right)
\lambda_3\left(\alpha,\beta\right).   \label{wet22}
\end{equation}
An analysis similar to that for Case 1 will show that the matrix elements
will satisfy the relations
\begin{eqnarray}
\frac{1}{2}\sum_{m_1m_3}
\left[\left\langle\lambda_1\mu_1 m_1;\lambda_3\mu_3m_3\left.
\right|\lambda_2\mu_2 m_2^{\prime}\right\rangle^{\ast}
\left\langle\Phi_{m_1}^{\lambda_1 \mu_1}\left|
T_{m_2}^{\lambda_2\mu_2} \right|\Phi_{m_3}^{\lambda_3\mu_3}
\right\rangle\right. ~~~~~&  & \nonumber \\
~~~~+ \left.\left\langle\lambda_1\mu_1 m_1;\lambda_3\mu_3m_3\left.
\right|\lambda_2\mu_2 m_2\right\rangle
\left\langle\Phi_{m_1}^{\lambda_1 \mu_1}\left|
 T_{m_2^{\prime}}^{\lambda_2\mu_2} \right|
\Phi_{m_3}^{\lambda_3\mu_3} \right\rangle 
\right]   &   &   \nonumber  \\
~~~~=~\frac{1}{d_{\lambda_2\mu_2}}\delta_{m_2,m_2^{\prime}}
\left\langle\lambda_1\mu_1\left|\left|\lambda_2\mu_2\right|
\right|\lambda_3 \mu_3\right\rangle_{AL},  &   &   \nonumber 
\end{eqnarray}
where the reduced matrix element for an antilinear operator $T$ is
\begin{equation}
\left\langle\lambda_1\mu_1\left|\left|\lambda_2\mu_2\right|\right|
\lambda_3\mu_3\right\rangle_{AL} = \sum_{n_1n_2n_3}\left\langle\lambda_1
\mu_1n_1;\lambda_3\mu_3n_3\left|\right.\lambda_2\mu_2n_2
\right\rangle^{\ast}\left\langle\Phi_{n_1}^{\lambda_1\mu_1}\left|
T_{n_2}^{\lambda_2\mu_2}\right|\Phi_{n_3}^{\lambda_3\mu_3}\right\rangle.
    \label{wet23}
\end{equation}
When none of these corepresentations are of type (b) according to
Wigner's classification \cite{wigner}, then
\begin{equation}
\left\langle\Phi_{m_1}^{\lambda_1\mu_1}\left|T_{m_2}^{\lambda_2\mu_2}
\right|\Phi_{m_3}^{\lambda_3\mu_3}\right\rangle = \frac{1}{d_{\lambda_2
\mu_2}}\left\langle\lambda_1 \mu_1\left|\left|\lambda_2\mu_2\right|\right|
\lambda_3\mu_3\right\rangle_{AL}\left\langle\lambda_1\mu_1 m_1;\lambda_3\mu_3
m_3\left|\right.\lambda_2\mu_2m_2\right\rangle,  \label{wet24}
\end{equation}
is a solution. The condition for essential uniqueness of this factorization
is given by a relation similar to Eq.~(\ref{wet21}) if the indices $2$
and $3$ there are interchanged. For linear groups again, the relation
(\ref{wet24}) is exact.

In case of reprtition of $D^{\lambda_2\mu_2}$ in the decomposition of the 
inner direct product representation $D^{\lambda_1\mu_1}\otimes
D^{\lambda_3\mu_3}$ the same considerations as in the case of linear $T$
will hold.  \\  \\

This work has been financed by the Department of Atomic Energy, India, \\
Project No. BRNS/Physics/14/74.  \\  \\
DEPARTMENT OF PHYSICS, UNIVERSITY OF KALYANI, \\
 KALYANI, WEST BENGAL, 741235 (INDIA)

\vspace{1.0cm}
{\sl P. M. van den Broek has kindly pointed out that even for 
corepresentations of type (b), the bases are orthogonal; hence the results 
obtained here are true for all the three types of corepresentations.} 
\end{document}